\documentclass[a4paper,twocolumn,11pt,accepted=2023-01-09]{quantumarticle}
\pdfoutput=1

\usepackage{graphicx}
\usepackage{dcolumn}
\usepackage{bm}
\usepackage[dvipsnames]{xcolor}
\usepackage{amsthm}
\usepackage{amsmath}
\usepackage{amssymb}
\usepackage[normalem]{ulem}
\usepackage{makecell} 
\usepackage{caption}
\usepackage{subcaption}
\usepackage{cite}

\newtheorem{obs}{Observation}

\newtheorem{defi}{Definition}
\newtheorem{prop}{Proposition}

\newtheorem{conje}{Conjecture}
\newtheorem{rem}{Remark}

\begin{document}


\title{Local hidden variable values without optimization procedures}
















\author{Dardo Goyeneche}
\affiliation{Departamento de F\'{i}sica, Facultad de Ciencias B\'{a}sicas, Universidad de Antofagasta, Casilla 170, Antofagasta, Chile}
\orcid{0000-0002-9865-4226}

\author{Wojciech Bruzda}
\affiliation{Institute of Theoretical Physics, Jagiellonian University, ul. {\L}ojasiewicza 11, 30-348 Krak\'ow, Poland}
\orcid{0000-0001-9743-7927}

\author{Ond\v{r}ej  Turek}
\affiliation{Department of Mathematics, Faculty of Science, University of Ostrava, 701 03 Ostrava, Czech Republic}
\orcid{0000-0002-5691-4907}

\author{Daniel Alsina}
\affiliation{School of Electronic and Electrical Engineering, University of Leeds, Leeds LS2 9JT, UK}
\orcid{0000-0002-2308-1750}

\author{Karol {\.Z}yczkowski}
\affiliation{Institute of Theoretical Physics, Jagiellonian University,  ul. {\L}ojasiewicza 11, 30-348 Krak\'ow, Poland}
\affiliation{Center for Theoretical Physics, Polish Academy of Sciences, Al.  Lotnik\'{o}w 32/46, 02-668 Warsaw, Poland}
\orcid{0000-0002-0653-3639}
\date{November 30, 2022}

\begin{abstract}
The problem of computing the local hidden variable (LHV) value of a Bell inequality plays a central role in the study of quantum nonlocality. In particular, this problem is the first step towards characterizing the LHV polytope of a given scenario. In this work, we establish a relation between the LHV value of bipartite Bell inequalities and the mathematical notion of excess of a matrix. Inspired by the well developed theory of excess, we derive several results that directly impact the field of quantum nonlocality. We show infinite families of bipartite Bell inequalities for which the LHV value can be computed exactly, without needing to solve any optimization problem, for any number of measurement settings. We also find tight Bell inequalities for a large number of measurement settings. 
\end{abstract}

\maketitle
\section{Introduction}
In a seminal paper \cite{B64}, John Bell proved 
that quantum correlations cannot be explained from deterministic and local hidden variable (LHV) models \cite{EPR35}. Since then,
an increasing interest in the field 
has triggered a large progress of the theory, supported by experimental implementations and practical applications. A feasible generation and certification of quantum nonlocality permits to design quantum technological applications having practical advantage with respect to its classical counterpart. For instance, quantum nonlocality can be used as a resource to outperform classical communication in certain distributed computing tasks \cite{CB97}, and enhances the communication power in the context of information theory \cite{CLMW11}. Quantum nonlocality has also led to the emergence of device-independent protocols, in which the involved parties do not need to trust their measurement devices \cite{L13,S13}. Concrete practical applications include quantum key distribution \cite{BHK05,AGM06,ABGMPS07,MPA11,PMLA13,VV14,KW16}, random number generation \cite{C07,CK11,PAMLMMOHLMM10,CR12}, quantum cryptography \cite{MPA11,PMLA13,VV14,KW16}, and testing quantumness of clouds of quantum computers \cite{AL16,GS18,SKPK19}, among others.

A fundamental open question in quantum nonlocality is the following: \emph{which Bell inequalities predict a quantum advantage with respect to LHV models?} Finding the answer to this question is crucial for the development of the practical applications shown above. For instance, a way to generate genuine random numbers \cite{HG17}, quantum cryptographic \cite{GRTZ02} and device independent \cite{B14} protocols rely on the nonlocal essence of Nature. In order to answer the aforementioned question, one needs to find both the LHV and quantum values. However, computing the LHV value of Bell inequalities  is  an  NP-hard  problem \cite{DBV17}. It is thus of fundamental relevance to search for a way to estimate the LHV value without involving any optimization procedure. In spite of a considerable effort made during
the last decades -- see \cite{BCPSW14} and references therein -- the complete understanding of the boundary 
between LHV theories and quantum nonlocality 
remains open even in bipartite scenarios composed by a small number of measurement settings and outcomes \cite{RBG14}.

In this work, we introduce a versatile tool that might shed some light to the above described problem. That is, we link the problem of finding the LHV value of bipartite Bell inequalities to the problem of finding matrices with maximal \emph{excess} \cite{S73,B77}. This one-to-one correspondence allows us to find the LHV value for infinitely many families of bipartite Bell inequalities in a wide range of scenarios. The observed connection plays an important role, specially when taking into account that identification of tight Bell inequalities relies on the study of LHV strategies.

It is important to remark here the relevance of our results and its comparison with further works existing in the literature. As the general problem of finding the LHV of Bell inequalities is too hard to be solved, there are two realistic approaches to the problem. The first one consists in defining an efficient algorithm to calculate the LHV of any given Bell inequality. The disadvantage here is that they efficiently work in relatively small scenarios only, see e.g. \cite{AHQ20}. The second approach consists in designing fast computational algorithms, or analytical procedures, to find the LHV in large scenarios, at the cost of restricting the attention to a subset of Bell inequalities, see e.g. \cite{PV22}. Thus, there is a trade-off between the efficiency of computing the LHV and the size of the set of Bell inequalities that one considers. The present work is more related to the second approach, producing some analytical results for Bell inequalities having any finite number of measurement settings and outcomes. Here, we emphasize that these results are explicitly given, without requiring to consider any computational algorithm.\medskip

This work is organized as follows. In Section \ref{sec:excess}, we  introduce our main tool, i.e. the excess of a matrix. In Section \ref{sec:bell}, we give a brief introduction to Bell inequalities. In Section \ref{sec:bell_excess}, we establish the one-to-one connection between maximal excess and the LHV value of bipartite Bell inequalities. This connection allows us to derive an upper bound for the LHV value that is stronger than known upper bounds for the quantum value. In Section \ref{sec:families}, we show how to find the LHV value of several classes of Bell inequalities without requiring to implement any optimization procedure. 
Additionally, in Section \ref{sec:tight} we find seven tight Bell inequalities for the two outcomes scenario with a large number of measurement settings, ranging between 8 and 20.  

\section{Excess of a matrix}\label{sec:excess} 
A square matrix $H$ with entries $\pm1$ is called {\sl Hadamard} if its columns are pairwise orthogonal. 
In 1973, Schmidt \cite{S73} asked about the maximal number of 1's that could be present in a Hadamard matrix of a given order $n$. 
To this end he introduced the notion of {\sl excess} of a Hadamard matrix $H$ as the difference between the number of the positive and negative entries of $H$. This value is usually denoted by $\Sigma(H)$. 
A derived quantity is the {\sl weight} of a Hadamard matrix $H$, defined as the number of $1$'s in $H$. Upper and lower bounds on the weight were first studied by Schmidt and Wang \cite{S77}.

 The maximal possible value of $\Sigma(H)$, among all Hadamard matrices of a given order $n$, called \emph{maximal excess}, was analyzed  by Best \cite{B77}, who found the following bounds: 
\begin{equation}\label{boundsE}
n^22^{-n}\binom{n}{\frac{n}{2}}\leq \max\Sigma(H)\leq n\sqrt{n}.
\end{equation}
where the maximum is taken over all existing Hadamard matrices of order $n$.

Moreover, there are infinitely many orders $n$ for which  $\max \Sigma(H)$ is already known \cite{B77,FK87,KF88,KS91,K91,XXS03}. For a Hadamard matrix of order $n$, upper bound (\ref{boundsE}) is attained
 if and only if $H$ is a \emph{constant row sum Hadamard matrix}, i.e. the sum of all entries in each row gives the same value \cite{B77}. Let us note that constant row sum matrices are sometimes called \emph{regular} in the context of Hadamard matrices \cite{B77}. 
Constant row sum Hadamard matrices of order $n$ only exist when $n$ is a square number. Thus, the upper bound (\ref{boundsE}) is not saturated when $n=2$. As we will see later, this has a direct connection with the fact that the CHSH Bell inequality \cite{CHSH69} has a quantum advantage.

Note that the excess of a Hadamard matrix is also equal to the sum of all entries of $H$. This allows a straightforward extension of the notion of excess to any complex square matrix $M$ of order $n$.
\begin{defi}
The sum of all entries of any square complex matrix $M$ of order $n$ is called its excess, denoted 
\begin{equation}\label{Sigma}
\Sigma(M)=\sum_{j,k=0}^{n-1}M_{jk}.
\end{equation}
\end{defi}
In Section \ref{sec:bell}, we restrict our attention to a class of matrices satisfying certain symmetries. In particular, any matrix within this class  has real excess and allows us to introduce the notion of maximal excess. In Section \ref{sec:bell_excess}, we will show that this class of matrices is one-to-one related to the full set of bipartite Bell inequalities.

\section{Bell inequalities}\label{sec:bell}
Suppose a bipartite scenario in which two observers, Alice and Bob, implement $m$ measurement settings per side having $q$ outcomes each. From an ensemble of identically prepared quantum states they can estimate a joint probability distribution $P(a,b|x,y)$, where  $a,b\in[0,\dots,q-1]$ denote outcomes for Alice and Bob, respectively, conditioned to the measurement settings $x,y\in[0,\dots,m-1]$, respectively. It can be shown that a single correlation of the form $P(a,b|x,y)$ does not provide enough evidence to ensure a conflict with LHV models \cite{F82}. However, a linear combination of such quantum probabilities attains values that cannot be reproduced by any LHV model \cite{B64}. Such expressions, known as \emph{Bell inequalities} \cite{BCPSW14}, are defined as follows:
\begin{equation}\label{Bell}
\sum_{x,y=0}^{m-1}\sum_{a,b=0}^{q-1}S^{ab}_{xy}\,P(a,b|x,y)\leq\mathcal{C},
\end{equation}
where $S^{ab}_{xy}$ is a real-valued function and $\mathcal{C}$, so called \emph{classical} or local hidden variable (LHV) value, is defined as the maximal achievable value of the left hand side in Eq.~(\ref{Bell}) in a local deterministic theory. That is, there is statistical independence between  Alice's and Bob's results, $P(a,b|x,y)=P(a|x)P(b|y)$, and the outcomes are deterministic, i.e. $P(a|x),P(b|y)\in\{0,1\}$, for every pair of measurement settings $x,y\in\{0,...,m-1\}$ and outcomes $a,b\in\{0,...,q-1\}$. The quantum value $\mathcal{Q}$ is defined as the maximal possible value of the left hand side in (\ref{Bell}), if optimization is implemented over all joint probability distributions admissible in quantum theory when observers implement local measurements and do not communicate their results. 
Probabilities in quantum theory take the form $P(a,b,|x,y)=\mathrm{Tr}[(\Pi_a^x\otimes\Pi_b^y) \rho_{AB}]$, where $\{\Pi_a^x\}$ and $\{\Pi_b^y\}$ define Positive-Operator Valued Measure (POVM), while $\rho_{AB}$ is a bipartite quantum state. 
The remarkable observation of Bell is that LHV correlations can be weaker than quantum correlations under certain conditions, thus it is possible to have $\mathcal{C}<\mathcal{Q}$. This important result, together with its experimental verification \cite{AGR81}, confirmed the nonlocal behavior of Nature.

Inequality (\ref{Bell}), when restricted to projective quantum measurements, can be represented by expectation values of correlators through the discrete double Fourier transform of the joint probability distribution  \cite{SATWAP17}:
\begin{equation}\label{FourierP}
P(a,b|x,y)=\frac{1}{q^2}\sum_{s,t=0}^{q-1}\omega^{as+bt} \langle A^s_x\otimes B^t_y\rangle,
\end{equation}
where $\omega=e^{2\pi i/q}$. Here, $A^s_x$ denotes the $s^{\rm th}$ power of the quantum observable $A_x$ associated to Alice, analogously for Bob. We assume that every observable has $q$ outcomes, associated with $q$ different roots of unity as eigenvalues. Note that any bipartite quantum probability distribution can be written as (\ref{FourierP}), even those produced by POVM measurements. This is due to the fact that any POVM can be seen as a projective measurement in a larger Hilbert space, due to the dilation theorem of Naimark \cite{N40}. 

Note that relation  (\ref{FourierP}) is constrained to quantum joint probability distributions, thus it agrees the normalization conditions
\begin{equation}
\sum_{a,b=0}^{q-1}P(a,b|x,y)=1
\end{equation}
for every $x,y=0,\dots,m-1$ and the no-signaling principle \cite{PR94}. That is,
\begin{equation}\label{nosignaling}
\begin{split}
\sum_{b=0}^{q-1}P(a,b|x,y)&=\sum_{b=0}^{q-1}P(a,b|x,y'),\\
\sum_{a=0}^{q-1}P(a,b|x,y)&=\sum_{a=0}^{q-1}P(a,b|x',y),
\end{split}
\end{equation}
for every $a,b=0,\dots, q-1$ and $x,y,x',y'=0,\dots,m-1$. Conditions (\ref{nosignaling}) imply that Alice and Bob cannot have faster-than-light communication. Equation (\ref{FourierP}) assumes quantum mechanics on the right hand side, agreeing with the no-signaling principle due to the fact that $A^0_x=B^0_y=\mathbb{I}$, for every $x,y=0,\dots,m-1$.

Due to (\ref{FourierP}), inequality  (\ref{Bell}) can be then rewritten as
\begin{equation}\label{BellM}
\sum_{x,y=0}^{m-1}\sum_{s,t=0}^{q-1}M_{ms+x,mt+y}\,\langle \\
A^{s}_{x}\otimes B^{t}_{y}\rangle\leq\mathcal{C},
\end{equation} 
where $M$ is a square matrix of order $n=mq$ with entries
\begin{equation}\label{matrix_M}
M_{ms+x,mt+y}=\frac{1}{q^2}\sum_{a,b=0}^{q-1}\omega^{sa+tb}\,S^{ab}_{xy},
\end{equation}
for every $0\leq s,t\leq q-1$, $0\leq x,y\leq m-1$. 
Entries of the matrix $M$ are defined as the Fourier transform (\ref{matrix_M}) of real valued functions. Thus, the following symmetric conditions hold: 
\begin{equation}\label{symm}
M_{m\left[q-s\right]_q+x,m\left[q-t\right]_q+y}=(M_{ms+x,mt+y})^*,
\end{equation}
for every $0\leq s,t\leq q-1$, $0\leq x,y\leq m-1$, where the asterisk denotes complex conjugation and $[x]_q$ denotes $x$ modulo $q$. 
The structure of matrices $M$ satisfying (\ref{symm}) is described in Appendix \ref{app:Structure_M}. 

The LHV value $\mathcal{C}$ of a bipartite Bell inequality can be alternatively calculated as follows, based on Eq. (\ref{BellM}):
\begin{equation}\label{LHVM}
\mathcal{C}=\max_{\substack{a_0,\dots,a_{m-1}\in\Omega_q\\b_0,\dots,b_{m-1}\in\Omega_q}} \sum_{x,y=0}^{m-1}\sum_{s,t=0}^{q-1}M_{ms+x,mt+y}\,
a^{s}_{x}\,b^{t}_{y},
\end{equation}
where $\Omega_q=\{1,\omega,\dots,\omega^{q-1}\}$. Eq.(\ref{LHVM}) plays a central role in our work, as we will see in Section \ref{sec:bell_excess}. 

\begin{rem}\label{remark_structureM}
Notice that if $q=2$, symmetry (\ref{symm}) implies that matrix $M$ is real. In general, any matrix $M\in\mathbb{C}^{mq\times mq}$ obeying condition \eqref{symm} has the following properties:
\begin{itemize}
    \item If $q$ is even, entries $M_{ms+x,mt+y}$ are real for all $x,y,s,t$ such that $0\leq x,y\leq m-1$ and $s,t\in\{0,q/2\}$.
    \item If $q$ is odd, entries $M_{x,y}$ are real for all \mbox{$0\leq x,y\leq m-1$}.
\end{itemize}
\end{rem}
In Appendix \ref{app:Structure_M}, we illustrate the structure of matrices $M$ for odd and even orders. 

We are now ready to introduce a notion of equivalence. We say that two Bell inequalities (\ref{Bell}) are equivalent 
if they differ in a relabeling of inputs and outputs only. That is, if their coefficients are given by $S^{ab}_{xy}$ and $S^{a'b'}_{x'y'}$, where  $S^{a'b'}_{x'y'}=\mathcal{P}S^{ab}_{xy}$, and $\mathcal{P}$ is a permutation. This notion of equivalence can be translated to matrices $M$ through (\ref{matrix_M}).
\begin{defi}\label{def:q_equivalence}
Two matrices $M$ and $M'$ of order $mq$ are $q$-equivalent
if there exists a permutation $\mathcal{P}^{a\to a',b\to b'}_{x\to x',y\to y'}$ such that  \begin{equation}
M'=\mathcal{F}\mathcal{P}\mathcal{F}^{-1}M,
\end{equation}
where $\mathcal{F}$ is the double Fourier transform defined in (\ref{matrix_M}).
\end{defi}

Definition \ref{def:q_equivalence} induces the following notion of maximal excess.
\begin{defi}\label{def:max_excess}
Let $q\geq2$. We say that $M$ of order $mq$ has maximal excess with respect to $q$-equivalence if $M$ obeys \eqref{symm} and
$$
\Sigma(M) = \max\{\Sigma(M');M'\text{ is $q$-equivalent to }M\}.
$$ 
\end{defi}
Note that any matrix $M'$, $q$-equivalent to $M$, satisfies the symmetric relation (\ref{symm}). This means that $\Sigma(M')$ is by construction a real quantity.

An important class of inequalities are the so-called \emph{correlation} Bell inequalities. These are inequalities of the form (\ref{BellM}) such that there are no marginal terms, i.e. $M_{ms+x,mt+y}=0$ when either $s=0$ or $t=0$. Physically, this means that every term of the Bell inequality (\ref{BellM}) is a linear combination of correlations existing between Alice and Bob. Let us introduce the entire set of matrices related to correlation Bell inequalities.
\begin{defi}
Let $q\geq2$. We say that a matrix $M$ is a correlation matrix if it satisfies (\ref{symm}) and  $M_{\alpha,\beta}=0$ when either  $\alpha=0$ or $\beta=0$. The nontrivial part of $M$, i.e. the submatrix of order $\mathfrak{n}=m(q-1)$ having entries $M_{\alpha,\beta}$ with $m\leq\alpha,\beta\leq mq-1$ is called the core of $M$.  \end{defi}
Note that the $q$-equivalence notion introduced in Definition \ref{def:q_equivalence} is equivalent to the standard notion of equivalence between two Hadamard matrices when $q=2$. Furthermore, Definition \ref{def:max_excess} reduces to the standard notion of the maximal excess for Hadamard matrices, introduced in Section \ref{sec:excess}, when $q=2$.

\section{Excess and LHV models}\label{sec:bell_excess}
In this section, we establish a connection between the notion of excess of certain class of matrices and the LHV value of a Bell inequality. The key observation is the following.  Given a classical value (\ref{LHVM}) for a matrix $M$, we can always find an equivalent matrix $M'$ such that  the optimal LHV strategy is achieved when all outputs are equal to $+1$, for both Alice and Bob. Thus, the LHV value of the related Bell inequality is given by the sum of all the entries of matrix $M'$, i.e. its excess. Let us now formalize this observation. 
\begin{prop}\label{excessLHV}
If $M$ is a matrix of order $mq$ having maximal excess with respect to $q$-equivalence, then the LHV value (\ref{LHVM}) equals the excess of the matrix $M$, i.e. $\mathcal{C}=\Sigma(M)$. Also, for any Bell inequality (\ref{BellM}) induced by a given matrix $M$, there is a $q$-equivalent matrix $M'$ such that $\mathcal{C}(M)=\Sigma(M')$.
\end{prop}

\begin{proof}
The optimal LHV strategy is given by all-one outcomes for both Alice and Bob. Thus, the LHV value is given by the sum of all the entries of matrix $M$. Conversely, if the  local variables producing the LHV value are $\{a_x\}$ and $\{b_y\}$, for Alice and Bob, respectively, then $M'_{[sx][ty]}=(a^s_xb^t_y)^*M_{[sx][ty]}$ produces an equivalent Bell inequality satisfying $\mathcal{C}(M')=\Sigma(M')$. Note that $M'$ satisfies the symmetry (\ref{symm}).
\end{proof}
The relevance of Prop.~\ref{excessLHV} consists in the fact that matrices $M$ having maximal excess have been exhaustively studied by mathematicians during the last five decades, including a wide range of Hadamard matrices \cite{B77,FK87,KF88,KS91,K91,XXS03,CK04}, complex Hadamard matrices \cite{KS93}, Hadamard tensors \cite{FG05} and orthogonal designs \cite{HLS78}. These results considerably extend the set of Bell inequalities for which the LHV value is known \cite{BCPSW14}, which is not a minor observation taking into account the hardness to calculate the LHV value of a bipartite Bell inequality \cite{ALMSS98}. 

For the rest of the work, we restrict our attention to matrices $M$ associated to Bell inequalities achieving the LHV value when the local deterministic strategies of both Alice and Bob have outcomes $+1$.

Let us illustrate Prop. \ref{excessLHV} with the celebrated CHSH inequality \cite{CHSH69}, associated to the matrix $$M=\left(\begin{array}{rrrr}
0&0&0&0\\
0&0&0&0\\
0&0&1&1\\
0&0&1&-1
\end{array}\right),
$$ of order $n=m\times q=2\times 2=4$, whose core is a Hadamard matrix of order $\mathfrak{n}=m\times (q-1)=2\times 1=2$:
\begin{equation}\label{CHSH}
\langle A_0\otimes B_0+A_0\otimes B_1+A_1\otimes B_0-A_1\otimes B_1\rangle\leq2.
\end{equation}
Here, $A_i$ and $B_j$ are dichotomic quantum observables represented by Hermitian operators having eigenvalues $\pm1$ each. This Hadamard matrix has maximal excess $\Sigma(M)=2$ \cite{B77}, coinciding with the LHV value of the CHSH inequality  \cite{CHSH69}. 

The next result requires the introduction of the numerical radius of a matrix $M$, $r(M)=\max_{|\psi\rangle\in\mathcal{H}_ n}|\langle\psi|M|\psi\rangle|$. Now, let us express an upper bound for excess in terms of the numerical radius.
\begin{prop}\label{Resupperr}
Let $M$ be a matrix of order $n=mq$ having maximal excess with respect to $q$-equivalence. Then, the following upper bound  holds for the LHV value $\mathcal{C}(M)$:
\begin{equation}\label{upperr}
\mathcal{C}(M)\leq nr(M).
\end{equation}
\end{prop}
\begin{proof}
Let $M$ be a matrix having maximal excess. The numerical radius is given by $r=\max_{|\psi\rangle\in\mathcal{H}}|\langle\psi|M|\psi\rangle|\geq\langle\phi|M|\phi\rangle$, where $|\phi\rangle=\frac{1}{\sqrt{n}}\sum_{i=0}^{n-1}|i\rangle=\frac{1}{\sqrt{n}}(1,1,\ldots,1)^T$. Thus $r\geq\langle\phi|M|\phi\rangle=\frac{1}{n}\Sigma(M)=\frac{1}{n}\mathcal{C}(M)$. 
\end{proof}

For any matrix $M$, we have that $\rho(M)\leq r(M)\leq\sigma(M)$, where $\rho(M)=\max_j|\lambda_j(M)|$ is the spectral radius  and $\sigma(M)=\max_j \sqrt{\lambda_j(MM^{\dag})}$ is the maximal singular value of $M$, also called spectral norm. For normal matrices, equalities $\rho(M)=r(M)=\sigma(M)$ hold. Consequently, the upper bound (\ref{upperr}) is more restrictive than a well-known upper bound for the quantum value, which we recall in the following Observation.
\begin{obs}[\cite{EKB13,V15}] The quantum value of a bipartite Bell inequality, induced by a matrix $M$ of order $n$, satisfies
\begin{equation}\label{upperQ}
\mathcal{Q}(M)\leq n\sigma(M).
\end{equation}
\end{obs}
The characterization of the entire set of matrices $M$ that saturate the bound (\ref{upperr}) or (\ref{upperQ}) is open in every scenario. Let us now provide a new upper bound for which we can fully characterize its tightness, in every bipartite scenario.
\begin{prop}\label{resUpperC}
Let $M$ be a matrix of order $n=mq$ having maximal excess with respect to $q$-equivalence. Then we have 
\begin{equation}\label{uppernu}
\mathcal{C}(M)\leq \sqrt{n}\,\nu(M),
\end{equation}
where $\nu(M)=\sqrt{\sum_{i=0}^{n-1}\bigl|\sum_{j=0}^{n-1}M_{ij}\bigr|^2}$. Moreover, upper bound (\ref{uppernu}) is saturated if and only if $M$ is a constant row sum matrix. In such case, $\mathcal{C}(M)=n\Gamma$, where $\Gamma$ is the constant row sum value.
\end{prop}
\begin{proof}
Upper bound (\ref{uppernu}) simply arises from the Cauchy-Schwarz inequality  
\begin{equation*}
\Sigma(M)=n\langle\phi|M|\phi\rangle\leq n\|M|\phi\rangle\|=n\frac{1}{\sqrt{n}}\,\nu(M).
\end{equation*}
Here, equality is attained if and only if $|\phi\rangle$ and $M|\phi\rangle$ are parallel, which occurs when each row of matrix $M$ has the same sum, i.e. $M$ has constant row sum. Also, note that $\mathcal{C}(M)=\Sigma(M)$, because $M$ has maximal excess.
\end{proof}
Notice that the constant row sum value $\Gamma$ occurring in Proposition \ref{resUpperC} is always a positive number due to the assumption that $M$ has maximal excess and to the symmetry assumption \eqref{symm}, implying that $\Sigma(M)\in\mathbb{R}$. Interestingly, inequality  (\ref{uppernu}) is saturated by all the matrices considered in nonlocal computation \cite{LPSW07} and quantum games having no quantum advantage \cite{RKMH14}. See Appendix \ref{app:example} for a particular example of analytic calculations of classical and quantum value for an infinite family of Bell inequalities.

Generalizing the above examples, note that Proposition \ref{resUpperC} provides the exact value of maximal excess for continuous families of matrices of any order. This allows to find the LHV for continuous families of Bell inequalities. For instance, when $q=2$, one can consider the family of real matrices $M$ having constant row sum $\Gamma$, such that this number coincides with the maximal eigenvalue of $M$. In this case, the LHV is equal to $n\Gamma$.

As a further example, consider the three inequivalent constant row sum Hadamard matrices of order $16$ \cite{H61}, defining three inequivalent correlation Bell inequalities in the bipartite scenario composed of $16$ measurement settings per side having $2$ outcomes each. These three inequalities have both LHV and quantum values equal to 64. The same conclusion does not apply to the Hadamard matrix $H_2$ of order 2, related to CHSH inequality \cite{CHSH69}, as this Hadamard matrix does not have constant row sum. 

On the other hand, non-unitary constant row sum matrices can imply a correlation Bell inequality having a quantum advantage. For instance, for $\mathfrak{n}=3$ the matrix
$\mathrm{core}(M)={\rm circ}\left[(0,-1, 1)\right]$
satisfies $\mathcal{C}(M)=4<3\sqrt{3}=\mathcal{Q}(M)$. Here, the quantum value $\mathcal{Q}(M)$ is achieved when considering measurement settings
$A_j$ and $B_k$ of the form $U_j D U^{\dagger}_j$, for
\begin{equation}
U_j=\left[\begin{array}{rr}
\cos\alpha_j & \sin\alpha_j\\
\sin\alpha_j  &  \cos\alpha_j
\end{array}\right],
\end{equation}
with $D=\{\{1,0\},\{0,-1\}\}$ and $\alpha_j=0$, $\frac{2\pi}{3}$, $\frac{\pi}{3}$ for $A_0$,
$A_1$, $A_2$ and $\alpha_j=\frac{\pi}{4}$, $\frac{7\pi}{12}$, $\frac{11\pi}{12}$ for $B_0$, $B_1$, $B_2$, respectively.

Note that a weaker version of the bound  (\ref{uppernu}) occurs when 
$\nu(M)=\|M\phi\|\leq\|M\|\|\phi\|=\sqrt{n}\sigma(M)$. In such case, the bound reduces to $\Sigma(M)\leq n\sigma(M)$, coinciding with the upper bound for the quantum value (\ref{upperQ}). From here and Prop. \ref{Resupperr}, an interesting fact arises.
\begin{obs}\label{circHconje}
The problem to classify all correlation Bell inequalities with $\mathfrak{n}$ settings for which there is no quantum advantage contains, as a special case, the problem of finding all constant row sum correlation matrices $M$ of order $\mathfrak{n}$, for which $\nu(M)$, defined in Prop. \ref{resUpperC}, equals the maximal singular value $\sigma(M)$. In particular, this problem contains the -long standing- circulant Hadamard conjecture \cite{R63}, which states that such matrices exist only for orders 1 and 4.
\end{obs}
There is a simple property that sometimes reduces the complexity of calculating excess of a matrix: excess of a tensor product is the product of excesses, see Observation 2.4 in \cite{OS07}. Applying this property to Bell inequalities we have the following result.
\begin{prop}\label{resprodcc}
Let $M_1$ and $M_2$ be matrices of order $n_1=m_1q$ and $n_2=m_2q$, respectively. If both $M_1$ and $M_2$ have maximal excess with respect to $q$-equivalence and $M_2$ is real, then the LHV value of the Bell inequality induced by the tensor product matrix $M=M_1\otimes M_2$, having $q$ outcomes per setting, is given by $\mathcal{C}(M)=\mathcal{C}(M_1)\mathcal{C}(M_2)$.
\end{prop}
\begin{proof}
The proof is straightforward from the fact that the excess of a tensor product of two matrices is equal to the product of excesses, see Obs. 2.4 in \cite{OS07}.
We only have to demonstrate that the matrix $M=M_1\otimes M_2$ obeys condition \eqref{symm}. By the definition of the tensor product, we have $[M_1\otimes M_2]_{ij}=[M_1]_{k_1l_1}\cdot[M_2]_{k_2l_2}$, where $i=k_1m_2q+k_2$, $j=l_1m_2+l_2$, $0\leq k_1,l_1<m_1q$, $0\leq k_2,l_2<m_2q$. Let $0\leq s,t<q$ and $0\leq x,y<m_1m_2q$. Writing $x=m_2qe_x+f_x$ and $y=m_2qe_y+f_y$ for $0\leq e_x,e_y<m_1$ and $0\leq f_x,f_y<m_2q$, we have
\begin{align*}
&[M_1\otimes M_2]_{m_1m_2q[q-s]_q+x,m_1m_2q[q-t]_q+y} \\
=&[M_1\otimes M_2]_{(m_1[q-s]_q+e_x)m_2q+f_x,(m_1[q-t]_q+e_y)m_2q+f_y} \\
=&[M_1]_{m_1[q-s]_q+e_x,m_1[q-t]_q+e_y}\cdot[M_2]_{f_x,f_y}\,,
\intertext{and since $M_1$ obeys \eqref{symm}, we have}
=&([M_1]_{m_1s+e_x,m_1t+e_y})^*\cdot[M_2]_{f_x,f_y}.
\end{align*}
Also, it holds
\begin{align*}
&([M_1\otimes M_2]_{m_1m_2qs+x,m_1m_2qt+y})^* \\
=&([M_1\otimes M_2]_{(m_1s+e_x)m_2q+f_x,(m_1t+e_y)m_2q+f_y})^* \\
=&([M_1]_{m_1s+e_x,m_1t+e_y}\cdot[M_2]_{f_x,f_y})^* \\
=&([M_1]_{m_1s+e_x,m_1t+e_y})^*\cdot([M_2]_{f_x,f_y})^*.
\end{align*}
Comparing these two expressions, we see that, if $M_2$ is real, then $M_1\otimes M_2$ obeys \eqref{symm}, and so $\Sigma(M)$ is real as well.
\end{proof}
Note that the product of matrices producing a tight Bell inequality does not induce a tight inequality, in general. Indeed, the inequality induced by core $H_4=H_2\otimes H_2$ is not tight, even thought CHSH inequality, induced by the core $H_2$, is tight. On the other hand, sometimes this occurs, as the core ${H_8=H_2\otimes H_2\otimes H_2}$ induces a tight Bell inequality, see Section \ref{sec:tight}.

\section{LHV value without optimization}\label{sec:families}
In this section, we show how to construct some families of correlation matrices $M$ of order $\mathfrak{n}=m(q-1)$ that allow us to achieve the LHV value of Bell inequalities with an unbounded number of measurement settings each, for $q=2$ outcomes. The following results summarize some relevant contributions to excess theory achieved in the last four decades. Before presenting the results, let us recall some required notions. A matrix $M$ is called skew symmetric if $M^T=-M$, where $T$ denotes transposition. A conference matrix is a square matrix having zero main diagonal, $\pm$1 off-diagonal entries and orthogonal columns 
A Hadamard matrix $H$ is skew-type if $H-\mathbb{I}$ is skew symmetric. Furthermore, recall that any real matrix $M$ satisfies (\ref{symm}), for $q=2$.
\begin{obs}[\cite{JKKS88}]\label{excessHadamardConference}
If $M$ is a skew-type Hadamard matrix of order $k\equiv0\,\mathrm{(mod\,4)}$ or a conference matrix of order $k\equiv 2\,(\mathrm{mod}\,4)$, then there is a Hadamard matrix $M$ of order $\mathfrak{n}=4k(k-1)$ (thus $\mathfrak{n}\equiv0\,(\mathrm{mod}\,16)$ and $\mathfrak{n}\equiv8\,(\mathrm{mod}\,16)$, respectively), with maximal excess
\begin{equation}\label{upperH}
\Sigma(M)= 4(k-1)^2(2k+1).
\end{equation}
\end{obs}
Note that none of the Hadamard matrices from Observation \ref{excessHadamardConference} are equivalent to one having constant row sum, as $\mathfrak{n}$ is not a square number. Thus, the related Bell inequalities are potential candidates to have a quantum advantage, in the sense that the upper bound (\ref{upperQ}) is strictly larger than the LHV value in each case. 

\begin{obs}[\cite{MS17}]
Let $M$ be a conference matrix of order $\mathfrak{n}$ and let $k$ be an odd integer such that ${k\leq\sqrt{\mathfrak{n}-1}<k+2}$. Then we have $$\Sigma(M)\leq\frac{\mathfrak{n}(k^2+2k+\mathfrak{n}-1)}{2(k+1)}.$$ Equality holds if and only if either (i) $\mathfrak{n}-1$ is a square and $M$ has constant row sum equal to $k$, or (ii) $\mathfrak{n}-1$ is not a  square and row sums are either $k$ or $k+2$.
\end{obs}

For instance, let us show an infinite family of Bell inequalities, arising from Hadamard matrices of order $\mathfrak{n}=\ell+1$, with $\ell\equiv 3\,(\mathrm{mod}\, 4)$, for which the maximal excess is known. The Hadamard matrices have the form 
\begin{equation}\label{Hffield}
M=\left(\begin{array}{cc}
-1&\mathbf{1}^T_{\ell}\\
\mathbf{1}_{\ell}&A
\end{array}\right),
\end{equation}
where $\mathbf{1}_{\ell}$ is the all-one vector of length $\ell$ and the square matrix $A$ of order $\ell$ is given by 
\begin{equation}
A_{ij}=\left\{\begin{array}{cl}
1&\mbox{if $j-i\in C\cup\{0\}$}\\
-1&\mbox{if $j-i\in \mathbb{F}_{\ell}\setminus(C\cup\{0\})$.}
\end{array}\right.
\end{equation}
Here, $\mathbb{F}_\ell$ denotes the finite field of order $\ell$ and $C$ is the set of nonzero squares of $\mathbb{F}_\ell$. This leads us to the following statement.
\begin{obs}[\cite{HS18}]
Let $m$ be an integer number such that $\ell=(2m+1)^2+2$ is the power of a prime, and $k$ be an even integer such that $k\leq\sqrt{\mathfrak{n}}<k+2$. Also, let $t=k$ if $|\mathfrak{n}-k^2|<|\mathfrak{n}-(k+2)^2|$ and $t=k-2$ otherwise. Then the Hadamard matrix (\ref{Hffield}) of order $\mathfrak{n}=\ell+1$ has maximal excess
\begin{equation}
\Sigma(M)=\mathfrak{n}(t+4)-4s,
\end{equation}
where $s$ is the integer part of the following expression: $\mathfrak{n}\bigl((t+4)^2-\mathfrak{n}\bigr)/(8t+16)$.
\end{obs}

As complementary information, a lower bound for the maximal excess of known Hadamard matrices has been found up to order $\mathfrak{n}=1000$, in many cases achieving the maximal excess \cite{JKKS88}. There are more explicit results for the maximal excess of Hadamard matrices than those shown above --- see  \cite{Y88,S89,KK90,KKS91} and references therein. The maximal excess has been also found for weaving Hadamard matrices \cite{CK04}, complex Hadamard matrices \cite{KS93} and orthogonal designs \cite{HLS78}. Furthermore, maximal excess for tensors \cite{FG05} opens the possibility to naturally extend our results to multipartite systems. Interestingly, maximal excess for tensors is connected with the so-called \emph{discrepancy of multivariable functions}, associated with certain multiparty communication problems, see \cite{FG05} and references therein. This connection might contribute to a better understanding of communication complexity for multipartite quantum systems \cite{BDHT99}.\medskip

We tend to believe the results and references shown in this section provide a valuable contribution to the theory of quantum nonlocality. On the one hand, this considerably extends the set of Bell inequalities for which the classical value is currently known \cite{BCPSW14}. On the other hand, the same refined techniques used in these references might inspire researchers to develop more efficient ways to find the LHV value for larger classes of Bell inequalities.

\begin{table}[ht!]
\begin{tabular}{c|c|c|c}
\thead{Number of \\ settings\\ $(m)$} &\thead{Number of\\ vertices}&\thead{Number of\\ affine inde-\\pendent\\ vertices}& \thead{Tightness}\\
 \hline
 2 & 4 & 3 & Tight\\ 
 \hspace{0.15cm}4* & 4 & 3 & Non-tight\\ 
 8 & 64 & 63&Tight$^{\dag}$\\ 
 12 &2640&143&Tight$^{\dag}$\\
 \hspace{0.15cm}16 (1)* &896&105&Non-tight\\
 \hspace{0.15cm}16 (2)* &192&81&Non-tight\\
 \hspace{0.15cm}16 (3)* &64&45&Non-tight\\
 16 (4) &21504&255&Tight$^{\dag}$\\
 16 (5) &21504&255&Tight$^{\dag}$\\
 20 (1)&20064&399&Tight$^{\dag}$\\
 20 (2)&20064&399&Tight$^{\dag}$\\
 20 (3)&20064&399&Tight$^{\dag}$\\
\end{tabular}
\caption{Study of tightness for bipartite correlation Bell inequalities of the form $\sum_{\alpha,\beta=0}^{\mathfrak{n}-1}H_{\alpha\beta}\langle A_{\alpha}\otimes B_{\beta}\rangle\leq\mathcal{C}$}, with $m=\mathfrak{n}$ measurement settings and $q=2$ outcomes. Here, $H$ is a Hadamard matrix of order  $\mathfrak{n}=m(q-1)$. Note that $H=\mathrm{core}(M)$ in (\ref{BellM}), where $M$ has order $n=mq$. All entries outside the core of $M$ are zero. The table shows the number of vertices of the LHV polytope that are touched by each hyperplane, and the number of affine independent vertices. For a tight Bell inequality, the latter number has to be equal to $n^2-1$, implying that the hyperplane forms a facet of the LHV polytope. The asterisk in the first column of the table means that the related Hadamard matrix has constant row sum. The seven tight cases denoted with symbol $\dag$ in the fourth column are new, to the best of our knowledge.
\label{tab:tight}
\end{table}
\section{Tight Bell inequalities}\label{sec:tight}
Tight Bell inequalities define facets of the LHV polytope. These hyperplanes completely characterize the set of correlations compatible with a LHV model. The quantum correlations space, for a bipartite scenario with $m$ settings and $q$ outcomes per party, is defined by all complex vectors  $v\in\mathbb{C}^{m^2q^2}$ having entries of the form $\langle A^s_x\otimes B^t_y\rangle$,    $x,y\in\{0,\dots,m-1\}$ and $s,t\in\{0,\dots,q-1\}$, where $A_x$ and $B_y$ are unitary operators having the $q^{\rm th}$ roots of unity as eigenvalues. The no-signaling conditions (\ref{nosignaling}) restrict quantum correlations to a $d=m^2(q-1)^2$ dimensional subspace, when single party measurements are not considered. Given that facets of the LHV polytope are faces with maximal dimension, they define hyperplanes in the correlation space. That is, they have dimension $d-1$. This means that for a two outcomes scenario ($q=2$) there are $d-1=m^2-1$ linearly independent vectors $v$, associated to LHV strategies defining a tight Bell inequality. 

Based on the above findings, in this section we study tightness of correlation Bell inequalities induced by Hadamard matrices $\mathrm{core}(M)$ up to order $\mathfrak{n}=20$. Results are summarized in Table \ref{tab:tight}, which leads us to establish the following conjecture:
\begin{conje}\label{conjecture}
A Hadamard matrix of order $\mathfrak{n}$ induces a tight correlation Bell inequality with $m=\mathfrak{n}$ settings and $q=2$ outcomes per party if and only if it is not equivalent to a matrix with constant row sum.
\end{conje}
Furthermore, in the study described in Table \ref{tab:tight}, we noted that the set of optimal LHV values can be arranged as \emph{mutually quasi-unbiased weighing matrices} \cite{NS15,KS18}, for both Alice and Bob strategies. See Appendix \ref{app:codes} for further details.

\section{Conclusions} We introduced a one-to-one relation between the local hidden variable (LHV) value of Bell inequalities and the mathematical notion of excess of a matrix, see Prop. \ref{excessLHV}. This connection allowed us to obtain -without any optimization- the LHV value of infinitely many families of Bell inequalities with an unbounded number of measurement settings per party, see Section \ref{sec:families}. 

We proposed the conjecture that every Hadamard matrix, not equivalent to a constant row sum one, induces a tight Bell inequality. We supported this conjecture with Hadamard matrices up to order 20, see Table \ref{tab:tight}.

Furthermore, we derived two upper bounds for the LHV value of bipartite Bell inequalities that are stronger than the upper bound for the quantum value, see Props. \ref{Resupperr} and \ref{resUpperC}. Tightness of the last bound has been fully characterized in every scenario, including some remarkable cases like all Bell inequalities related to nonlocal computation \cite{LPSW07} and quantum {\sc xor} games without quantum advantage \cite{RKMH14}.


Finally, we have shown that the problem to characterize the set of bipartite Bell inequalities with no quantum advantage for two outcomes contains the circulant Hadamard conjecture, a long-standing open problem in Combinatorics, see Observation \ref{circHconje}.

Our results provide new insights by merging two extensively studied areas of research coming from different fields. Quantum nonlocality seems to be the most benefited side, as the mathematical theory of excess is considerably more advanced than currently known techniques to find local hidden variable values, as far as we know. We believe our results will deepen the interest in quantum nonlocality among both physicists and mathematicians.
 \bigskip

\emph{Acknowledgements} Authors kindly acknowledge valuable discussions with A. Ac\'{i}n, R. Augusiak, A. Cabello, J. Calsamiglia, D. Cavalcanti, R. Craigen, J. De Vicente, P.~Horodecki, C. Jebaratnam, 
R.~Ramanathan, O. Reardon-Smith, G.~Senno, J. Tura and A.~Winter. D.G. is supported Grant FONDECYT Iniciaci\'{o}n number 11180474, Chile. This work was partially done by D.G. during his visit to the Jagiellonian University, Krakow, Poland, supported by MINEDUC-UA project, code ANT 1999. W.B. and K.{\.Z}. are supported by 
Narodowe Centrum Nauki under the Maestro grant number DEC-2015/18/A/ST2/00274,
by Foundation for Polish Science under the Team-Net
project no. POIR.04.04.00-00-17C1/18-00. D.A. is supported by UK EPSRC Grant No. EP/M013472/1.

\appendix

\section{Structure of matrices M}\label{app:Structure_M}

There are two structure for matrices $M$: \emph{(i)} when $q$ is odd, the top-left $m\times m$ square block is real. For instance, for $q=5$ and $m=3$ we have a matrix of the form
\begin{scriptsize}
\begin{equation*}
\left(
\begin{array}{ccccccccccccccc}
 \text{R} & \text{R} & \text{R} & . & . & . & . & . & . & . & . & . & . & . & . \\
 \text{R} & \text{R} & \text{R} & . & . & . & . & . & . & . & . & . & . & . & . \\
 \text{R} & \text{R} & \text{R} & . & . & . & . & . & . & . & . & . & . & . & . \\
 . & . & . & . & . & . & . & . & . & . & . & . & . & . & . \\
 . & . & . & . & . & . & . & . & . & . & . & . & . & . & . \\
 . & . & . & . & . & . & . & . & . & . & . & . & . & . & . \\
 . & . & . & . & . & . & . & . & . & . & . & . & . & . & . \\
 . & . & . & . & . & . & . & . & . & . & . & . & . & . & . \\
 . & . & . & . & . & . & . & . & . & . & . & . & . & . & . \\
 . & . & . & . & . & . & . & . & . & . & . & . & . & . & . \\
 . & . & . & . & . & . & . & . & . & . & . & . & . & . & . \\
 . & . & . & . & . & . & . & . & . & . & . & . & . & . & . \\
 . & . & . & . & . & . & . & . & . & . & . & . & . & . & . \\
 . & . & . & . & . & . & . & . & . & . & . & . & . & . & . \\
 . & . & . & . & . & . & . & . & . & . & . & . & . & . & . \\
\end{array}
\right).
\end{equation*}
\end{scriptsize}
Here, letter $R$ denotes real entries and dots complex entries satisfying (\ref{symm}).
\emph{(ii)} when $q$ is even, there are 4 $m\times m$ real blocks, associated to $s,t\in\{0,q/2\}$, for $0\leq x,y\leq m-1$. For instance, for $q=4$ and $m=3$ we have a matrix of the form
\begin{scriptsize}
\begin{equation*}
\left(
\begin{array}{cccccccccccc}
 \text{R} & \text{R} & \text{R} & . & . & . & \text{R} & \text{R} & \text{R} & . & . & . \\
 \text{R} & \text{R} & \text{R} & . & . & . & \text{R} & \text{R} & \text{R} & . & . & . \\
 \text{R} & \text{R} & \text{R} & . & . & . & \text{R} & \text{R} & \text{R} & . & . & . \\
 . & . & . & . & . & . & . & . & . & . & . & . \\
 . & . & . & . & . & . & . & . & . & . & . & . \\
 . & . & . & . & . & . & . & . & . & . & . & . \\
 \text{R} & \text{R} & \text{R} & . & . & . & \text{R} & \text{R} & \text{R} & . & . & . \\
 \text{R} & \text{R} & \text{R} & . & . & . & \text{R} & \text{R} & \text{R} & . & . & . \\
 \text{R} & \text{R} & \text{R} & . & . & . & \text{R} & \text{R} & \text{R} & . & . & . \\
 . & . & . & . & . & . & . & . & . & . & . & . \\
 . & . & . & . & . & . & . & . & . & . & . & . \\
 . & . & . & . & . & . & . & . & . & . & . & . \\
\end{array}
\right),
\end{equation*}
\end{scriptsize}
where, again, $R$ denotes real entries and dots represent complex entries satisfying (\ref{symm}).


\section{Optimal LHV strategies}
\label{app:codes}

This Appendix is devoted to showing that optimal LHV strategies for both Alice and Bob, when arranged as rows of matrices, define Mutually quasi-unbiased weighing matrices, for some correlation Bell inequalities arising from Hadamard matrices of orders $\mathfrak{n}=2$ and $\mathfrak{n}=8$. Each of these matrices is the core of the matrix $M$ that generates the corresponding correlation Bell inequality, in the scenario of two parties, $m=\mathfrak{n}$ settings and 2 outcomes each.

A weighing matrix $W$ of weight $k$ is a square matrix of order $m$ having entries from the set $\{-1,0,1\}$, such that $WW^T=k\mathbb{I}$. Two weighing matrices $W_1$ and $W_2$ of order $m$ and weight $k$ are \emph{mutually quasi-unbiased weighing matrices} (MQUWM) \cite{KS18} for parameters $(m,k,l,a)$ if there exists positive integers $a$ and $l$ such that $(1/\sqrt{a})W_1W^T_2$ is a weighing matrix of weight $l=k^2/a$. We denote these matrices as MQUWM$(m,k,l,a)$.\medskip

\textbf{Case} $\mathbf{m=2}$ \textbf{settings (CHSH):}\smallskip

There are 4 optimal LHV strategies in this case. For Alice and Bob, the strategies define the following two sets of two MQUWM(2,2,1,4):
\begin{equation}
W^A_1=\left(\begin{array}{rr}
1&1\\
1&-1
\end{array}\right)\hspace{0.5cm}W^A_2=\left(\begin{array}{rr}
1&-1\\
1&1
\end{array}\right),
\end{equation}
and\begin{equation}
W^B_1=\left(\begin{array}{rr}
1&1\\
-1&1
\end{array}\right)\hspace{0.5cm}W^B_2=\left(\begin{array}{rr}
1&-1\\
1&1
\end{array}\right),
\end{equation}
respectively. That is, we have two orthonormal bases in each case, all of them identical up to global phases and reordering of vectors.\medskip

\textbf{Case} $\mathbf{m=4}$ \textbf{settings:}\medskip

There are 4 optimal LHV strategies. For both Alice and Bob, the strategies can be arranged in rows as the following weighing matrices of weight 4, i.e. Hadamard matrices:
\begin{equation}
W^A_1=\left(\begin{array}{rrrr}
1&-1&-1&-1\\
1&-1&1&1\\
1&1&-1&1\\
1&1&1&-1
\end{array}\right),
\end{equation}
and
\begin{equation}
W^B_1=\left(\begin{array}{rrrr}
-1&1&1&1\\
1&-1&1&1\\
1&1&-1&1\\
1&1&1&-1
\end{array}\right),
\end{equation}
respectively. In this case, we also have two orthonormal bases. \medskip

\textbf{Case} $\mathbf{m=8}$ \textbf{settings:}\medskip

There are 64 optimal LHV strategies. For Alice, her local strategies can be arranged as the following maximal set of 8 MQUWM(8,8,16,4), along the rows:
\small{
\begin{equation*}
W^A_1=\left(
\begin{array}{rrrrrrrr}
 1 & -1 & -1 & -1 & -1 & -1 & -1 & -1 \\
 1 & -1 & -1 & -1 & 1 & 1 & 1 & 1 \\
 1 & -1 & 1 & 1 & -1 & -1 & 1 & 1 \\
 1 & -1 & 1 & 1 & 1 & 1 & -1 & -1 \\
 1 & 1 & -1 & 1 & -1 & 1 & -1 & 1 \\
 1 & 1 & -1 & 1 & 1 & -1 & 1 & -1 \\
 1 & 1 & 1 & -1 & -1 & 1 & 1 & -1 \\
 1 & 1 & 1 & -1 & 1 & -1 & -1 & 1 \\
\end{array}
\right)
\end{equation*}

\begin{equation*}
W^A_2=\left(
\begin{array}{rrrrrrrr}
 1 & -1 & -1 & -1 & -1 & -1 & 1 & 1 \\
 1 & -1 & -1 & -1 & 1 & 1 & -1 & -1 \\
 1 & -1 & 1 & 1 & -1 & -1 & -1 & -1 \\
 1 & -1 & 1 & 1 & 1 & 1 & 1 & 1 \\
 1 & 1 & -1 & 1 & -1 & 1 & 1 & -1 \\
 1 & 1 & -1 & 1 & 1 & -1 & -1 & 1 \\
 1 & 1 & 1 & -1 & -1 & 1 & -1 & 1 \\
 1 & 1 & 1 & -1 & 1 & -1 & 1 & -1 \\
\end{array}
\right)
\end{equation*}

\begin{equation*}
W^A_3=\left(
\begin{array}{rrrrrrrr}
 1 & -1 & -1 & -1 & -1 & 1 & -1 & 1 \\
 1 & -1 & -1 & -1 & 1 & -1 & 1 & -1 \\
 1 & -1 & 1 & 1 & -1 & 1 & 1 & -1 \\
 1 & -1 & 1 & 1 & 1 & -1 & -1 & 1 \\
 1 & 1 & -1 & 1 & -1 & -1 & -1 & -1 \\
 1 & 1 & -1 & 1 & 1 & 1 & 1 & 1 \\
 1 & 1 & 1 & -1 & -1 & -1 & 1 & 1 \\
 1 & 1 & 1 & -1 & 1 & 1 & -1 & -1 \\
\end{array}
\right)
\end{equation*}

\begin{equation*}
W^A_4=\left(
\begin{array}{rrrrrrrr}
 1 & -1 & -1 & -1 & -1 & 1 & 1 & -1 \\
 1 & -1 & -1 & -1 & 1 & -1 & -1 & 1 \\
 1 & -1 & 1 & 1 & -1 & 1 & -1 & 1 \\
 1 & -1 & 1 & 1 & 1 & -1 & 1 & -1 \\
 1 & 1 & -1 & 1 & -1 & -1 & 1 & 1 \\
 1 & 1 & -1 & 1 & 1 & 1 & -1 & -1 \\
 1 & 1 & 1 & -1 & -1 & -1 & -1 & -1 \\
 1 & 1 & 1 & -1 & 1 & 1 & 1 & 1 \\
\end{array}
\right)
\end{equation*}

\begin{equation*}
W^A_5=\left(
\begin{array}{rrrrrrrr}
 1 & -1 & -1 & 1 & -1 & -1 & -1 & 1 \\
 1 & -1 & -1 & 1 & 1 & 1 & 1 & -1 \\
 1 & -1 & 1 & -1 & -1 & -1 & 1 & -1 \\
 1 & -1 & 1 & -1 & 1 & 1 & -1 & 1 \\
 1 & 1 & -1 & -1 & -1 & 1 & -1 & -1 \\
 1 & 1 & -1 & -1 & 1 & -1 & 1 & 1 \\
 1 & 1 & 1 & 1 & -1 & 1 & 1 & 1 \\
 1 & 1 & 1 & 1 & 1 & -1 & -1 & -1 \\
\end{array}
\right)
\end{equation*}

\begin{equation*}
W^A_6=\left(
\begin{array}{rrrrrrrr}
 1 & -1 & -1 & 1 & -1 & -1 & 1 & -1 \\
 1 & -1 & -1 & 1 & 1 & 1 & -1 & 1 \\
 1 & -1 & 1 & -1 & -1 & -1 & -1 & 1 \\
 1 & -1 & 1 & -1 & 1 & 1 & 1 & -1 \\
 1 & 1 & -1 & -1 & -1 & 1 & 1 & 1 \\
 1 & 1 & -1 & -1 & 1 & -1 & -1 & -1 \\
 1 & 1 & 1 & 1 & -1 & 1 & -1 & -1 \\
 1 & 1 & 1 & 1 & 1 & -1 & 1 & 1 \\
\end{array}
\right)
\end{equation*}

\begin{equation*}
W^A_7=\left(
\begin{array}{rrrrrrrr}
 1 & -1 & -1 & 1 & -1 & 1 & -1 & -1 \\
 1 & -1 & -1 & 1 & 1 & -1 & 1 & 1 \\
 1 & -1 & 1 & -1 & -1 & 1 & 1 & 1 \\
 1 & -1 & 1 & -1 & 1 & -1 & -1 & -1 \\
 1 & 1 & -1 & -1 & -1 & -1 & -1 & 1 \\
 1 & 1 & -1 & -1 & 1 & 1 & 1 & -1 \\
 1 & 1 & 1 & 1 & -1 & -1 & 1 & -1 \\
 1 & 1 & 1 & 1 & 1 & 1 & -1 & 1 \\
\end{array}
\right)
\end{equation*}

\begin{equation*}
W^A_8=\left(
\begin{array}{rrrrrrrr}
 1 & -1 & -1 & 1 & -1 & 1 & 1 & 1 \\
 1 & -1 & -1 & 1 & 1 & -1 & -1 & -1 \\
 1 & -1 & 1 & -1 & -1 & 1 & -1 & -1 \\
 1 & -1 & 1 & -1 & 1 & -1 & 1 & 1 \\
 1 & 1 & -1 & -1 & -1 & -1 & 1 & -1 \\
 1 & 1 & -1 & -1 & 1 & 1 & -1 & 1 \\
 1 & 1 & 1 & 1 & -1 & -1 & -1 & 1 \\
 1 & 1 & 1 & 1 & 1 & 1 & 1 & -1 \\
\end{array}
\right).
\end{equation*}
}
For Bob, we also have a maximal set of 8 MQUWM(8,8,16,4), with a different ordering of vectors:
\small{
\begin{equation*}
W^B_1=\left(
\begin{array}{rrrrrrrr}
 -1 & 1 & 1 & 1 & 1 & 1 & 1 & 1 \\
 1 & -1 & -1 & 1 & -1 & 1 & 1 & 1 \\
 1 & 1 & 1 & 1 & -1 & -1 & -1 & 1 \\
 1 & 1 & -1 & -1 & 1 & 1 & -1 & 1 \\
 1 & -1 & 1 & -1 & 1 & -1 & 1 & 1 \\
 1 & 1 & 1 & -1 & -1 & 1 & 1 & -1 \\
 1 & 1 & -1 & 1 & 1 & -1 & 1 & -1 \\
 1 & -1 & 1 & 1 & 1 & 1 & -1 & -1 \\
\end{array}
\right)
\end{equation*}
\begin{equation*}
W^B_2=\left(
\begin{array}{rrrrrrrr}
 -1 & 1 & -1 & 1 & -1 & 1 & 1 & 1 \\
 -1 & -1 & 1 & 1 & 1 & 1 & -1 & 1 \\
 1 & 1 & 1 & 1 & -1 & 1 & -1 & -1 \\
 1 & 1 & -1 & 1 & 1 & -1 & -1 & 1 \\
 -1 & 1 & 1 & 1 & 1 & -1 & 1 & -1 \\
 1 & -1 & 1 & 1 & -1 & -1 & 1 & 1 \\
 1 & 1 & 1 & -1 & 1 & 1 & 1 & 1 \\
 1 & -1 & -1 & 1 & 1 & 1 & 1 & -1 \\
\end{array}
\right)
\end{equation*}
\begin{equation*}
W^B_3=\left(
\begin{array}{rrrrrrrr}
 -1 & -1 & 1 & 1 & -1 & 1 & 1 & 1 \\
 -1 & 1 & -1 & 1 & 1 & -1 & 1 & 1 \\
 -1 & 1 & 1 & -1 & 1 & 1 & -1 & 1 \\
 1 & 1 & -1 & 1 & -1 & 1 & -1 & 1 \\
 1 & 1 & 1 & -1 & -1 & -1 & 1 & 1 \\
 1 & -1 & 1 & 1 & 1 & -1 & -1 & 1 \\
 1 & 1 & 1 & 1 & 1 & 1 & 1 & -1 \\
 1 & -1 & -1 & -1 & 1 & 1 & 1 & 1 \\
\end{array}
\right)
\end{equation*}
\begin{equation*}
W^B_4=\left(
\begin{array}{rrrrrrrr}
 -1 & 1 & 1 & -1 & -1 & 1 & 1 & 1 \\
 1 & -1 & 1 & 1 & -1 & 1 & -1 & 1 \\
 -1 & 1 & 1 & 1 & 1 & 1 & -1 & -1 \\
 1 & 1 & -1 & 1 & 1 & 1 & 1 & 1 \\
 1 & 1 & 1 & 1 & -1 & -1 & 1 & -1 \\
 -1 & -1 & 1 & 1 & 1 & -1 & 1 & 1 \\
 1 & -1 & 1 & -1 & 1 & 1 & 1 & -1 \\
 1 & 1 & 1 & -1 & 1 & -1 & -1 & 1 \\
\end{array}
\right)
\end{equation*}
\begin{equation*}
W^B_5=\left(
\begin{array}{rrrrrrrr}
 -1 & 1 & 1 & 1 & -1 & 1 & 1 & -1 \\
 -1 & -1 & -1 & 1 & 1 & 1 & 1 & 1 \\
 1 & 1 & -1 & 1 & -1 & -1 & 1 & 1 \\
 1 & 1 & 1 & 1 & 1 & 1 & -1 & 1 \\
 -1 & 1 & 1 & -1 & 1 & -1 & 1 & 1 \\
 1 & -1 & 1 & -1 & -1 & 1 & 1 & 1 \\
 1 & -1 & 1 & 1 & 1 & -1 & 1 & -1 \\
 1 & 1 & -1 & -1 & 1 & 1 & 1 & -1 \\
\end{array}
\right)
\end{equation*}
\begin{equation*}
W^B_6=\left(
\begin{array}{rrrrrrrr}
 -1 & 1 & 1 & 1 & -1 & -1 & 1 & 1 \\
 1 & 1 & -1 & 1 & -1 & 1 & 1 & -1 \\
 1 & 1 & 1 & -1 & -1 & 1 & -1 & 1 \\
 -1 & 1 & -1 & 1 & 1 & 1 & -1 & 1 \\
 1 & -1 & 1 & 1 & 1 & 1 & 1 & 1 \\
 -1 & 1 & 1 & -1 & 1 & 1 & 1 & -1 \\
 1 & 1 & -1 & -1 & 1 & -1 & 1 & 1 \\
 1 & 1 & 1 & 1 & 1 & -1 & -1 & -1 \\
\end{array}
\right)
\end{equation*}
\begin{equation*}
W^B_7=\left(
\begin{array}{rrrrrrrr}
 -1 & 1 & 1 & 1 & -1 & 1 & -1 & 1 \\
 -1 & 1 & -1 & 1 & 1 & 1 & 1 & -1 \\
 1 & 1 & -1 & -1 & -1 & 1 & 1 & 1 \\
 1 & -1 & -1 & 1 & 1 & 1 & -1 & 1 \\
 -1 & -1 & 1 & -1 & 1 & 1 & 1 & 1 \\
 1 & -1 & 1 & 1 & -1 & 1 & 1 & -1 \\
 1 & 1 & 1 & 1 & 1 & -1 & 1 & 1 \\
 1 & 1 & 1 & -1 & 1 & 1 & -1 & -1 \\
\end{array}
\right)
\end{equation*}
\begin{equation*}
W^B_8=\left(
\begin{array}{rrrrrrrr}
 1 & 1 & 1 & 1 & -1 & 1 & 1 & 1 \\
 -1 & 1 & 1 & 1 & 1 & -1 & -1 & 1 \\
 -1 & 1 & -1 & -1 & 1 & 1 & 1 & 1 \\
 1 & 1 & -1 & 1 & 1 & 1 & -1 & -1 \\
 -1 & -1 & 1 & 1 & 1 & 1 & 1 & -1 \\
 1 & -1 & -1 & 1 & 1 & -1 & 1 & 1 \\
 1 & 1 & 1 & -1 & 1 & -1 & 1 & -1 \\
 1 & -1 & 1 & -1 & 1 & 1 & -1 & 1 \\
\end{array}
\right)
\end{equation*}
}
Let us now see how these optimal LHV strategies of Alice and Bob are connected. The $i^{\rm th}$ element of the $j^{\rm th}$ basis $W^A_j$ is connected with the $k^{\rm th}$ element of the $l^{\rm th}$ basis $W^B_l$ through the following ordering $(i,j,k,l)$: 
\small{
\begin{eqnarray*}
&(1,1,1,1),(2,1,1,8),(3,1,4,4),(4,1,4,5),(5,1,5,6),&\\
&(6,1,7,7),(7,1,7,2),(8,1,7,3),(1,2,1,2),(2,2,1,7),&\\
&(3,2,4,6),(4,2,4,3),(5,2,5,1),(6,2,7,5),(7,2,7,4),&\\
&(8,2,7,8),(1,3,1,3),(2,3,1,6),(3,3,4,1),(4,3,4,8),&\\
&(5,3,6,4),(6,3,6,2),(7,3,8,5),(8,3,8,7),(1,4,1,4),&\\
&(2,4,1,5),(3,4,4,7),(4,4,4,2),(5,4,6,8),(6,4,6,3),&\\
&(7,4,6,6),(8,4,6,1),(1,5,2,5),(2,5,3,1),(3,5,3,8),&\\
&(4,5,3,2),(5,5,5,7),(6,5,5,4),(7,5,8,3),(8,5,8,6),&\\
&(1,6,2,3),(2,6,2,4),(3,6,2,7),(4,6,3,6),(5,6,6,5),&\\
&(6,6,5,2),(7,6,8,8),(8,6,7,1),(1,7,2,2),(2,7,3,5),&\\
&(3,7,3,7),(4,7,3,4),(5,7,5,8),(6,7,5,3),(7,7,7,6),&\\
&(8,7,8,1),(1,8,2,1),(2,8,2,8),(3,8,3,3),(4,8,2,6),&\\
&(5,8,5,5),(6,8,6,7),(7,8,8,2),(8,8,8,4).&
\end{eqnarray*}
}

Regarding higher order Hadamard matrices, we studied tightness without analyzing their geometrical structure, due to the large number of classical strategies achieving the LHV value.\medskip

\textbf{Case} $\mathbf{m=12}$ \textbf{settings:}\medskip

For order 12, there is an unique Hadamard matrix, see the catalog of Hadamard matrices by J. Seberry~\cite{JSCatalogue}. In this case, we find a tight Bell inequality, composed by 12 measurement settings per party and 2 outcomes each.

\textbf{Case} $\mathbf{m=16}$ \textbf{settings:}\medskip

We noted that the three constant row sum Hadamard matrices of order 16, denoted $H_1, H_2$ and $H_3$ \cite{JSCatalogue}, do not imply tight Bell inequalities, for two outcomes. This is consistent with the fact that correlation tight Bell inequalities always have a quantum violation \cite{ECW19}. The remaining two non-constant row sum Hadamard matrices, $H_4$ and $H_5$, lead to a quantum violation and produce inequivalent tight Bell inequalities, composed by 16 measurement settings per party and 2 outcomes each.\medskip

\textbf{Case} $\mathbf{m=20}$ \textbf{settings:}\medskip

 For order $m=20$, we found that the three Hadamard matrices $H_1$, $H_2$, $H_3$ \cite{JSCatalogue} induce inequivalent tight Bell inequalities, composed by 20 measurement settings per party and 2 outcomes each.\medskip

The above analysis include the entire set of classes of Hadamard matrices existing up to order $m=20$. Recall that two equivalent Hadamard matrices imply essentially the same Bell inequality, up to a relabeling of inputs and outputs. See Table \ref{tab:tight} and Conjecture \ref{conjecture} for a summary of these results.

\section{Bell inequalities for any $q\geqslant 2$}\label{app:example}

In this section, we illustrate our method by presenting an infinite family of Bell inequalities with $m=q$ settings and $q$ outcomes, for which we find its LHV for any $q\geq2$. The family is characterized by the following matrix of order $q^2$:
\begin{equation}\label{Fsquare}
M_{ms+x,mt+y}=\omega^{xt-sy},
\end{equation}
where $0\leq s,t\leq q$, $0\leq x,y\leq m$ and $\omega=e^{2\pi i/q}$. This hermitian version of the Fourier matrix -only existing for square orders- has been used to study complex equiangular tight frames \cite{BE10,Sz13,GT17}. The first observation is that (\ref{Fsquare}) naturally satisfies the required symmetry (\ref{symm}), thus implying a Bell inequality (\ref{BellM}).

Let us now show that this family of Bell inequalities satisfy $\mathcal{C}=q^3$, for every $q\geq2$. To do so, let $a_x=\omega^x$ and $b_y=\omega^{-y}$ be LHV strategies for Alice and Bob, respectively, where $0\leqslant x,y\leqslant m-1$. Thus, we have \begin{equation*}
\sum_{x,y,s,t=0}^{q-1} F_{qs+x,qt+y}a_x^sb_y^t=\sum_{x,y,s,t=0}^{q-1} \omega^{x(s+t)-y(s+t)}=q^3,
\end{equation*}
which saturates the upper bound (\ref{uppernu}). Therefore, the chosen LHV strategy is a solution to the optimization problem (\ref{LHVM}), implying that $\mathcal{C}=q^3$.

On the one hand, the value $q^3$ is also the upper bound for the quantum value (\ref{upperQ}), implying that this family of Bell inequalities does not have quantum advantage for any $q\geq2$. On the other hand, by using the inverse of (\ref{matrix_M}) it is simple to see that these Bell inequalities coincide with the bipartite version of the \emph{Guess Your Neighbour Input game} \cite{ABBAGP10}:
\begin{equation}\label{GYNI}
\sum_{a,b=0}^{q-1}P(a,b|b,a)\leq q.
\end{equation}
Inequalities (\ref{GYNI}) satisfy $\mathcal{C}=\mathcal{Q}=\mathcal{NS}$ for every $q\geq2$, where $\mathcal{NS}$ is the maximal value allowed in no-signaling theories. This implies that these inequalities are not physically interesting, in the sense that they do not define a border between two kind of correlations. Nonetheless, it is interesting the fact that these inequalities are related to an hermitian version of the Fourier transform. We leave as an open problem to find further Butson type complex Hadamard matrices, inequivalent to the Fourier matrix of order $q^2$, such that they satisfy the symmetry (\ref{symm}). Finally, note that any real Hadamard matrix of order $mq$ satisfies the symmetry (\ref{symm}), for $q=2$.

\end{document}